\documentclass[12pt]{spieman}   
\usepackage[]{graphicx}
\usepackage{setspace}
\usepackage{tocloft}

\title{The capabilities and performance of the Automated Planet Finder Telescope with the implementation of a dynamic scheduler} 

\author{Jennifer Burt\supscr{a}, Bradford Holden\supscr{a}, Russell Hanson\supscr{a}, Greg Laughlin\supscr{a}, Steve Vogt\supscr{a}, Paul Butler\supscr{b}, Sandy Keiser\supscr{b}, William Deich\supscr{a}}

\affiliation{\supscrsm{a}UCO/Lick Observatory, 1156 High Street, Santa Cruz, CA, USA, 95060\\
\supscr{b} Department of Terrestrial Magnetism, Carnegie Institution of Washington, Washington, DC, USA, 20015}

\cftpagenumbersoff{figure}
\cftpagenumbersoff{table} 
\begin{document} 
\maketitle 

\newcommand{\ms}{${\rm m~s^{-1}}$}
\newcommand{\degs}{${\rm ^{\circ}~s^{-1}}$}

\begin{abstract}

We report initial performance results emerging from 600 hours of observations with the Automated Planet Finder (APF) telescope and Levy Spectrometer located at UCO/Lick Observatory. We have obtained multiple spectra of 80 G, K and M-type stars, which comprise 4,954 individual Doppler radial velocity (RV) measurements with a median internal uncertainty of 1.35 \ms.

We find a strong, expected correlation between the number of photons accumulated in the 5000-6200\AA\ iodine region of the spectrum, and the resulting internal uncertainty estimates. Additionally, we find an offset between the population of G and K stars and the M stars within the data set when comparing these parameters. As a consequence of their increased spectral line densities, M-type stars permit the same level of internal uncertainty with 2x fewer photons than  G-type and K-type stars. When observing M stars, we show that the APF/Levy has essentially the same speed-on-sky as Keck/HIRES for precision RVs.

In the interest of using the APF for long-duration RV surveys, we have designed and implemented a dynamic scheduling algorithm. We discuss the operation of the scheduler, which monitors ambient conditions and combines on-sky information with a database of survey targets to make intelligent, real-time targeting decisions. 
 
\end{abstract}


{\noindent \footnotesize{\bf Address all correspondence to}: Jennifer Burt, UCO/Lick Observatory, 1156 High St, Santa Cruz, USA, 95060; E-mail:  \linkable{jaburt@ucolick.edu} }


\section{Introduction}
\label{sect:intro}

The Doppler Velocity technique now has two decades of success in enabling extrasolar planetary detections. It has produced candidates with masses approaching that of Earth, and has been especially successful in detecting long period planets. In recent years, Doppler Velocity confirmations have proven vital to gaining an understanding of the planetary candidates discovered by photometric space missions, especially Kepler. Doppler Velocity campaigns are responsible for 31\% of the 1854 confirmed planetary discoveries in the past three decades, but they account for 87\% of the 330 confirmed planets with periods longer than 1 year\cite{ExoplanetsEU}. Ground-based facilities, furthermore, are amenable to the operation of long-term surveys due to their relativity low construction and operational costs along with their ability to be upgraded as instrumentation improves.

In order to find analogs of our own solar system, we need to extend the catalog of successful radial velocity (RV) planet detections to encompass longer-period planets (particularly true Jupiter analogs) and smaller mass, short-period planets. This means observing efforts must increase their temporal baselines and cadence of observation to more effectively populate each planet's RV phase curve. 

The Automated Planet Finder, located at the Mt. Hamilton station of UCO/Lick Observatory, combines a 2.4-m telescope with a purpose-built, high-resolution echelle spectrograph, and is capable of 1\ms\ Doppler Velocity precision\cite{Vogt2014}. Eighty percent of the telescope's observing time is specifically dedicated to the detection of extrasolar planets. This time is shared evenly between two exoplanet research groups, one at UC Santa Cruz and one at UC Berkeley. Time is allocated in whole night segments, with a schedule developed quarterly by the telescope manager. Target lists and operational software are developed separately as the two exoplanet groups are focused on different types of planet detection/follow up. For a description of the UC Berkeley planet detection efforts, please see Fulton et. al, 2015\cite{Fulton2015}. The remaining 20\% of telescope time is dedicated to at-large use by the University of California community. All users are allowed to request specific nights if it is beneficial to their science goals (for example: to obtain RV values while a planet transits its star), and such requests are taken into account by the telescope manager when setting the schedule.

The APF leverages a number of inherent advantages to improve efficiency. For example, its Levy spectrometer, a high-resolution prism cross-dispersed echelle spectrograph with a maximum spectral resolving power of R$\sim150,000$, is optimized for high precision RV planet detection \cite{Radovan2010,Vogt2014}. A full description of the design and the individual components of the APF is available in Vogt et al. (2014) \cite{Vogt2014}. 

To support long-running surveys, we have developed a dynamic scheduler capable of making real-time observing decisions and running the telescope without human interaction. Through automation and optimization, we increase observing efficiency, decrease operating costs and minimize the potential for human error. The scheduler's target selection is driven by balancing scientific goals (what we {\it want} to observe based on scientific interest, required data quality, and desired cadences) and engineering constraints (what we {\it can} observe based on current atmospheric conditions and physical limitations of the telescope). To address these criteria, we need to know how the velocity precision extracted from a given stellar spectrum depends on inputs that can be monitored before and during each observation. We have assessed the influences of the various inputs by analyzing 16 months of data taken on the APF between June 2013 \& October 2014.

The plan of this paper is as follows: in \S\,\ref{sect:data} we describe the current APF radial velocity catalog, paying special attention to the variety of spectral types and the frequency of observations. In \S\,\ref{sect:Inputs} we evaluate the relations between velocity precision and parameters including stellar color and V band magnitude, airmass, seeing, date of observation and atmospheric transparency and we explain how these relations inform the nightly decision-making process executed by the observing software.  In \S\,\ref{sect:DismissedInputs} we outline the parameters that were assumed to be important prior to on-sky observations, but that have since been determined to have little relevance. In \S\,\ref{sect:Heimdallr} we describe how the relevant relations are integrated into the scheduling software, and we discuss its structure, its dependencies, and its capabilities. In \S\, \ref{sect:compare} we discuss other automated and semi-automated observatories and highlight both the similarities and differences between those systems and that employed by the APF. Finally we conclude in \S\,\ref{sect:campaigns} by reviewing the application of the APF's automated observing strategy to the telescope's current and upcoming observing scientific campaigns. 

\section{Data set description}
\label{sect:data}

\subsection{Description of observing terminology}

\begin{figure}[h]
   \begin{center}
   \includegraphics[height=8cm]{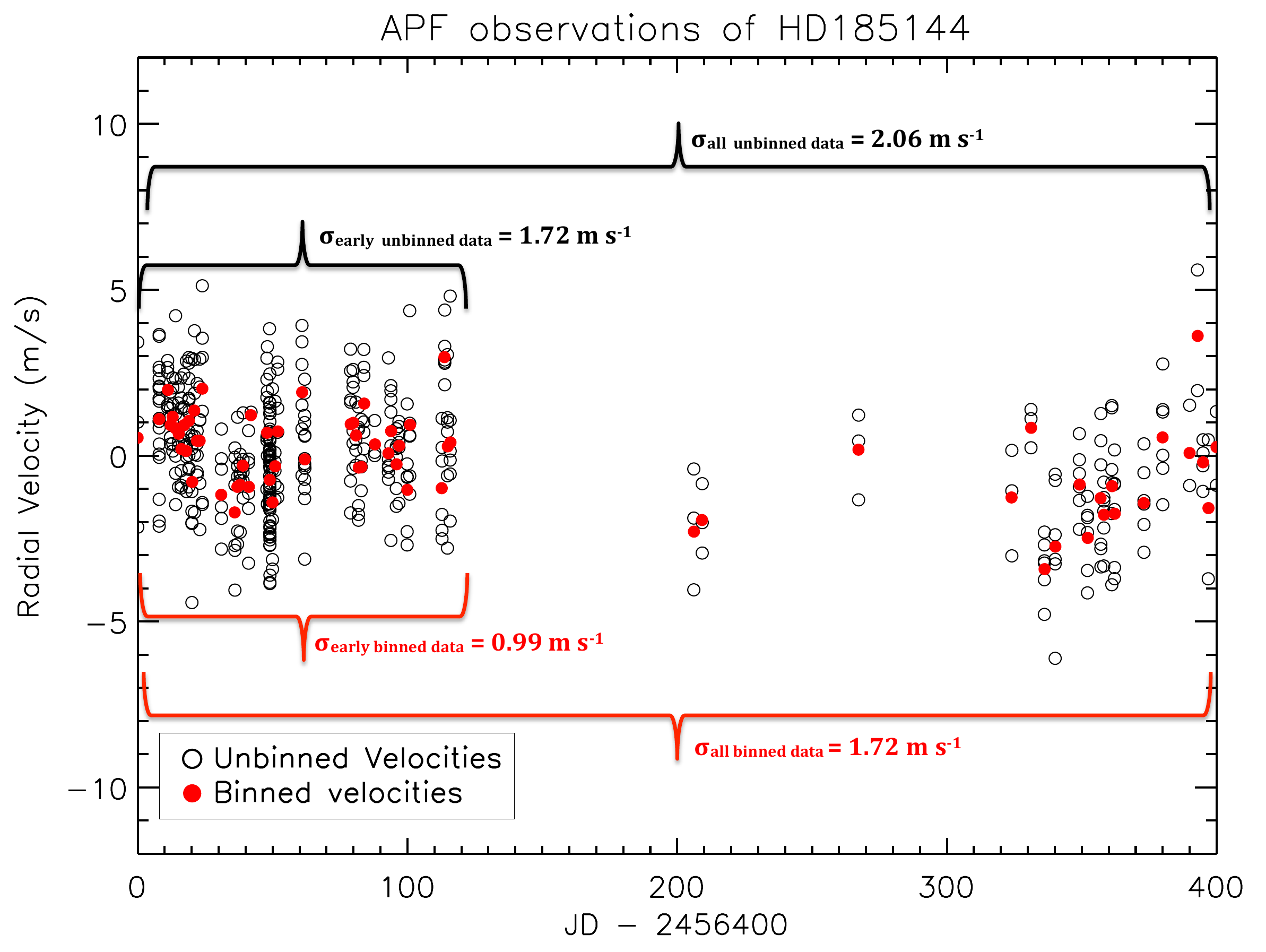}  
   \end{center}
   \caption 
   { \label{fig:bin_unbin} Our radial velocity dataset for the RV standard star HD 185144 showing the individual exposures (open circles in black) and the resulting binned velocities (solid circles in red). The binned velocities have a much smaller RMS value (0.99 \ms\ for the first 120 days and 1.72 \ms\ over the entire 400 day span, compared to the 1.72 \ms\ and 2.06 \ms\ exhibited by the un-binned data over the same respective time spans) due to their increased signal-to-noise. Additionally the individual exposures are each $\sim$60 seconds long while the binned velocities span at least $\sim$5min, so they average over the p-modes of the star. This paper uses the individual exposure data when carrying out all further analyses and calibrations. The change in the RMS values between the first 120 days and the full data set is largely due to a change in observing strategy. After the first 4 months we started observing HD 185144 less frequently and with fewer exposures in each observation to allow for more targets to be observed each night.} 
   \end{figure}

This paper differs from most publications discussing precision radial velocity work in that all of the plots, equations and discussions presented are based on individual, un-binned exposures of stars. It is well known that pulsation modes (p-modes) in stars cause oscillations on the stellar surface, adding noise to the RV signal. It has also been well documented that the noise imparted by these p-modes in late-type stars can be averaged over by requiring total observation times longer than the $\sim 5-15$ minute periods typical of the pulsation cycles \cite{Santos2004, Dumusque2011}. Thus the majority of radial velocity publications, especially those dealing with exoplanet detections, present {\it binned} velocities and error values. That is, they take multiple, individual {\it exposures} of a star during the night and then combine (bin) them to create one final {\it observation} with its own velocity and internal error estimate \cite{Dumusque2011, Burt2014, Fulton2015}. The binned observations therefore contain more photons than any individual exposure, but, more importantly, average over the pulsation modes on the star, and therefore exhibit a measurably smaller scatter (Figure \ref{fig:bin_unbin}).

These terms, exposure and observation, will be used specifically throughout this paper to help make clear to the reader whether we are talking about an exposure - a single instance of the shutter being open and collecting photons from the star, or an observation - the combination of all exposures taken of a star over the course of a night. Similarly, exposure time will refer to the open shutter time during one exposure and observation time will refer to the total open shutter time spent on a target between all exposures. In the case of the star's observation consisting of only one exposure, then the observation time is equivalent to the exposure time.

\subsection{The APF data set}

The APF's star list is made up of legacy targets first observed as part of the Lick-Carnegie exoplanet campaign using the HIRES spectrograph on the Keck I telescope. Stars for the Lick-Carnegie survey were selected based on three main criteria: 
\begin {itemize}
\item Spectral type between G and M : stars with spectral types earlier than F5 experience oscillations that can produce pseudo-Keplerian signals, and are fast rotators with few spectral lines so they are avoided
\item Single stars : no stars with companions closer than 5", as this can lead to scattered light from the nearby companion making its way into the spectrograph slit and contaminating the target star's signal
\item Quiet stars : Only stars without large amounts of emission in the Ca II H \& K line cores are permitted, as core emission in these lines is an indicator of stellar chromospheric activity \cite{Noyes1984} which complicates the data reduction process and can produce pseudo-Keplerian signals.
\end{itemize}

The resulting Lick-Carnegie target list is comprised of $\sim$1800 stars, which have been monitored using Keck/HIRES over the past two decades. When creating the initial target list for the APF, the Lick-Carnegie star list was culled for targets with V magnitudes brighter than 12 and declinations above -20$^{\circ}$. In order to efficiently prove the APF's capabilities, we selected stars with suspected short-period planets (P \textless 100 days) that required only 1-2 more rounds of phase coverage to verify. The presence and the false-alarm probabilities of these short-period, Keplerian signals were determined by analyzing the existing Keck/HIRES RV datasets using the publicly available Systemic Console \cite{Meschiari2009}. Systemic allows users to fit planetary signals to RV data and derive the orbital properties, while also providing tools to handle error estimation and assess orbital stability. This selection process resulted in a list of 127 stars.

\begin{figure}[h]
   \begin{center}
   \includegraphics[height=8cm]{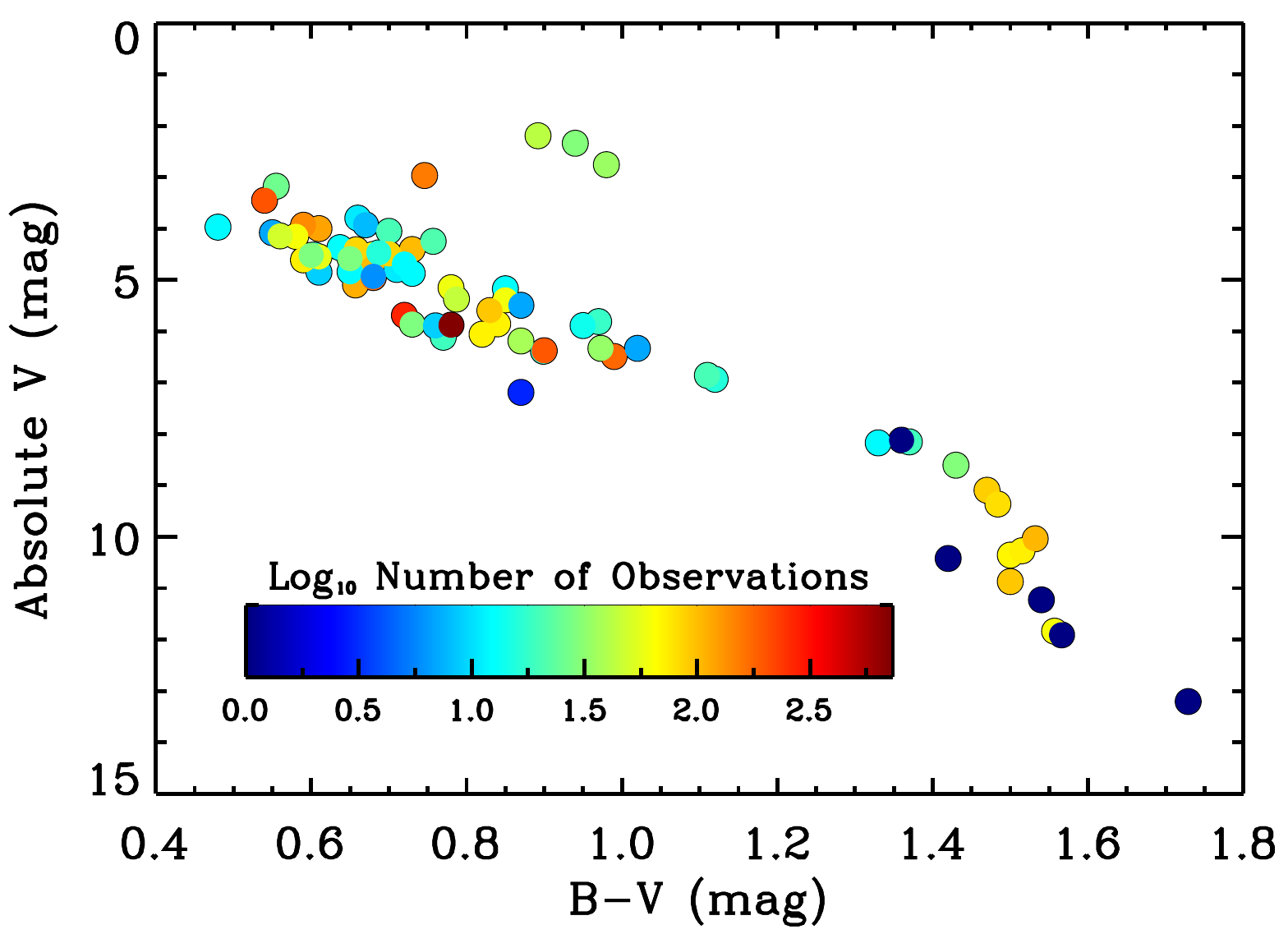}  
   \end{center}
   \caption 
   { \label{fig:APF_CMD} Color-Magnitude diagram for stars observed with APF and used in the analysis described herein. Color coding represents the number of exposures that have been obtained for each star. Our dataset spans a wide range in color and magnitude and contains stars with spectral types from F6 to M4.} 
   \end{figure}

The calibrations described in this paper are based on data taken with the APF between June 2013 and October 2014. The data set includes precision Doppler observations of 80 of the 127 stars selected from the Lick-Carnegie survey and this is before the development/inclusion of the dynamic scheduler, thus all data in this paper were obtained using fixed star lists. Our precision Doppler observations encompass spectra of 80 stars and incorporate 600 hours of open shutter time. The stars span spectral types from early G to mid M, have 3.5 \textless V \textless 12, and are all located within 160 pc (Figure \ref{fig:APF_CMD}). 

Every APF star has a set of observations containing between one and seven hundred exposures. Individual exposures are restricted to a maximum length of fifteen minutes to avoid cosmic ray accumulation and to minimize uncertainty when calculating the photon-weighted midpoint times. Additionally, we enforce a total observing time limit of one hour per target per night to ensure that telescope time is not wasted on observing faint objects when conditions are poor. 

For each individual exposure, the $\rm \overline{FWHM}$ (the average FWHM of the star over the integration time) of the guide camera's seeing disk is logged in the FITS image header, along with the total number of photons from the exposure meter and the total exposure time. Colors, magnitudes and distances for the stars are obtained from SIMBAD \cite{Wegner2000}.

All spectra are analyzed using the data reduction techniques described in detail by Butler et al. (1996)\cite{Butler1996}, which produces a measurement of the stellar RVs and associated internal uncertainties. The reduction pipeline analyzes each exposure's extracted spectrum in 2\AA\ chunks and determines the radial velocity shift for each chunk individually. The final reported velocity is the average over all the chunks, while the internal uncertainty is defined as the RMS of the individual chunk velocity values about the mean, divided by the square root of the number of chunks. Thus the internal uncertainty represents errors in the fitting process, which are dominated by the photon noise. The internal uncertainty does not include potential systematic errors associated with the instrument, nor does it account for astrophysical noise (or "jitter") associated with the star and, therefore, represents only a lower limit to the accuracy of the data for finding companions.

\subsection{Determining additional systematic errors}

We use the APF's 737 individual exposures of HD 185144 (Sigma Draconis, HR 7462), a bright RV standard star, to estimate the precision with which we can measure radial velocities. It should be noted that the radial velocity values produced by our analysis pipeline are all relative velocities and thus have a mean value of zero. We examine the unbinned, or individual, exposures, finding that star's mean internal uncertainty is $\mu_{int} = 1.0$ \ms  and that the mean of the absolute values of its velocity measurements is $\mu_{abs} = 1.8$ \ms. Because $\mu_{abs}$ includes effects from the internal uncertainty {\it in addition} to other sources of error such as the stellar jitter and the instrument systematics, its value is always higher than the internal uncertainty value for a given exposure.

\begin{figure}[h]
   \begin{center}
   \includegraphics[height=8cm]{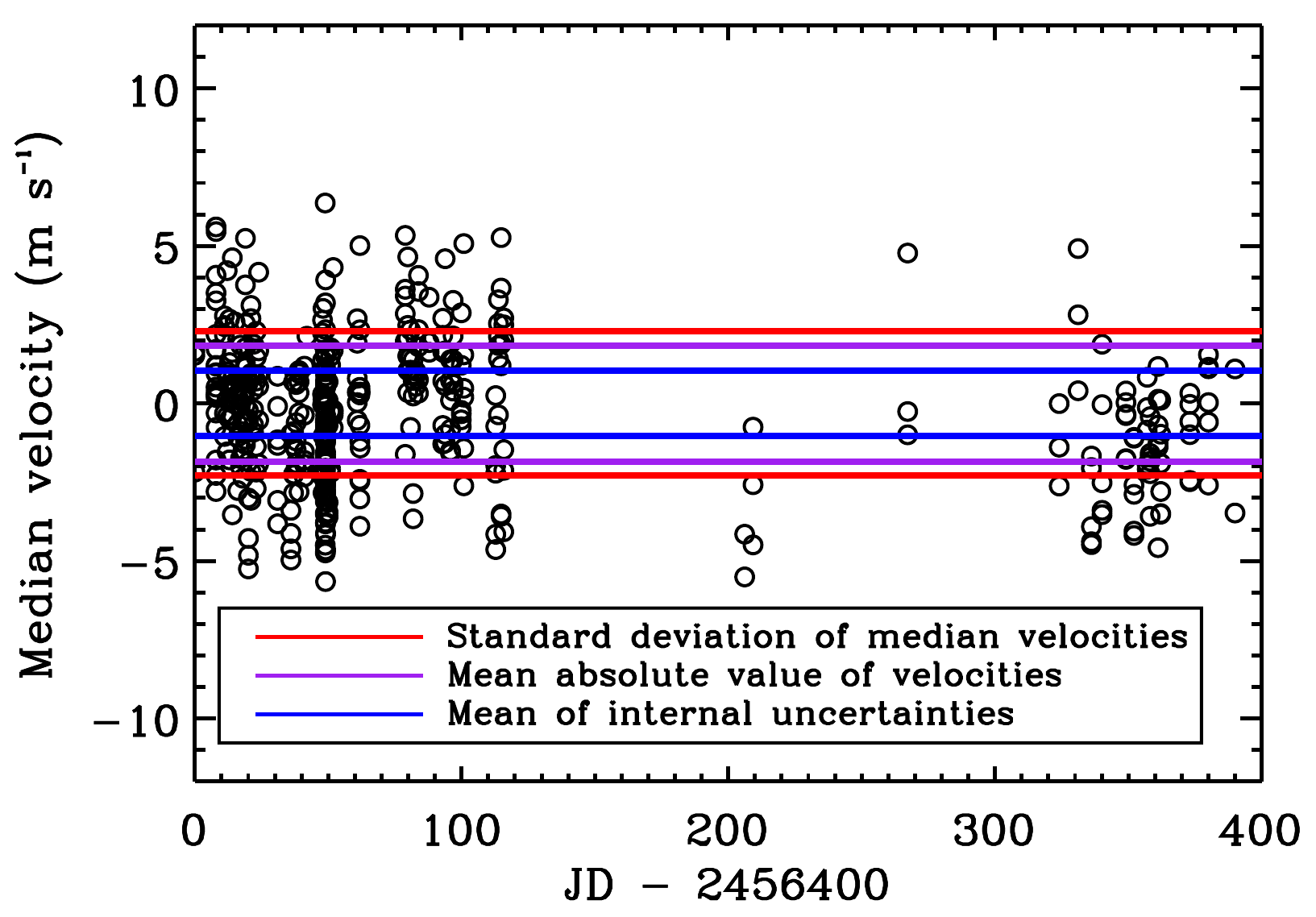}  
   \end{center}
   \caption 
   { \label{fig:185144_unc} Radial velocities from APF exposures on the star HD 185144 [G9V, $ \rm m_{v}$ = 4.68], 737 points in total. The internal uncertainty estimates produced by the data reduction pipeline are noticeably smaller than the actual spread in the data. The internal error does not account for telescope systematic errors or sources of astrophysical noise in the star. Our average internal uncertainty is $\mu_{int}$ = 1.0 \ms (blue line) but we find the mean absolute value of the HD185144 velocity measurements to be $\mu_{abs}$ = 1.8 \ms (purple line). We adopt an estimate that the additional systematic uncertainty in our velocity precision is $\sigma$ = 1.5 \ms. } 
   \end{figure}

The difference between these two parameters implies an additional quadrature offset of $1.5$ \ms, which we then generalize as $\sigma_{s} = 1.5$ \ms (Figure \ref{fig:185144_unc}). We take this offset value to represent the additional error contributions from all other systematics, including the known 5-15 minute pulsation modes of the star that the binned data averages over and the systematics from the instrument. If, instead, we compared the mean internal uncertainty to the standard deviation of the velocity measurements ($\sigma_{vel} = 2.3$ \ms) we would require a larger additional offset term ($\sigma_{s} = 2.0$ \ms)\footnote{As stated, the internal uncertainties and velocities used in this analysis are extracted from the individual spectra obtained by the telescope. Our normal operational mode determines the internal uncertainties from data binned on a two hour time scale, which is the approach used by Vogt et al. (2014)\cite{Vogt2014} and yields a standard deviation of $\sigma_{bin} = 1.05$ \ms\ for HD 185144. This value is notably smaller than the standard deviation of the individual velocities, $\sigma_{vel} = 2.3$ \ms, because we deliberately acquired 6 observations of HD 185144 in order to both average over the pulsation modes and achieve a high precision for the final binned observations.}.  We note, however, that the data are strongly affected by the velocities of exposures that fall in the outlying, non-Gaussian tail. We thus choose to determine our error estimate using the mean of the absolute values of the velocity measurements (instead of the standard deviation of the velocities) as it mitigates the influence of the outliers. 

We use the relations of Wright et al. (2005)\cite{Wright2005}, which presents radial velocity jitter estimates at the 20$^{th}$ percentile, median and 80$^{th}$ percentile levels to assess the expected stellar activity for HD 185144. We find an estimated median jitter of $\sigma'_{rv} = 3.5$ \ms, and a 20$^{th}$ percentile value of $2.3$ \ms. Noting that the $20^{th}$ percentile value matches the $\sigma_{vel}$ for our exposures, obtained over a timespan of 400 days, suggests that the star is in fact intrinsically quiet. That is to say, we expect only 20\% of stars with the same evolution metric, activity metric (both described by Wright 2005) and B-V color as HD 185144 to have activity levels less than $2.3$ \ms. Given that we measure $\mu_{abs} = 1.8$ \ms, HD 185144 is, at a minimum, much quieter than expected based on its color, activity and evolution. Even so, it is reasonable to expect that some fraction of the observed uncertainty is due to the star itself (astrophysical noise) or currently unknown planets rather than instrumental effects, thus indicating that the instrument is performing very well.

\begin{figure}[h]
   \begin{center}
   \includegraphics[height=8cm]{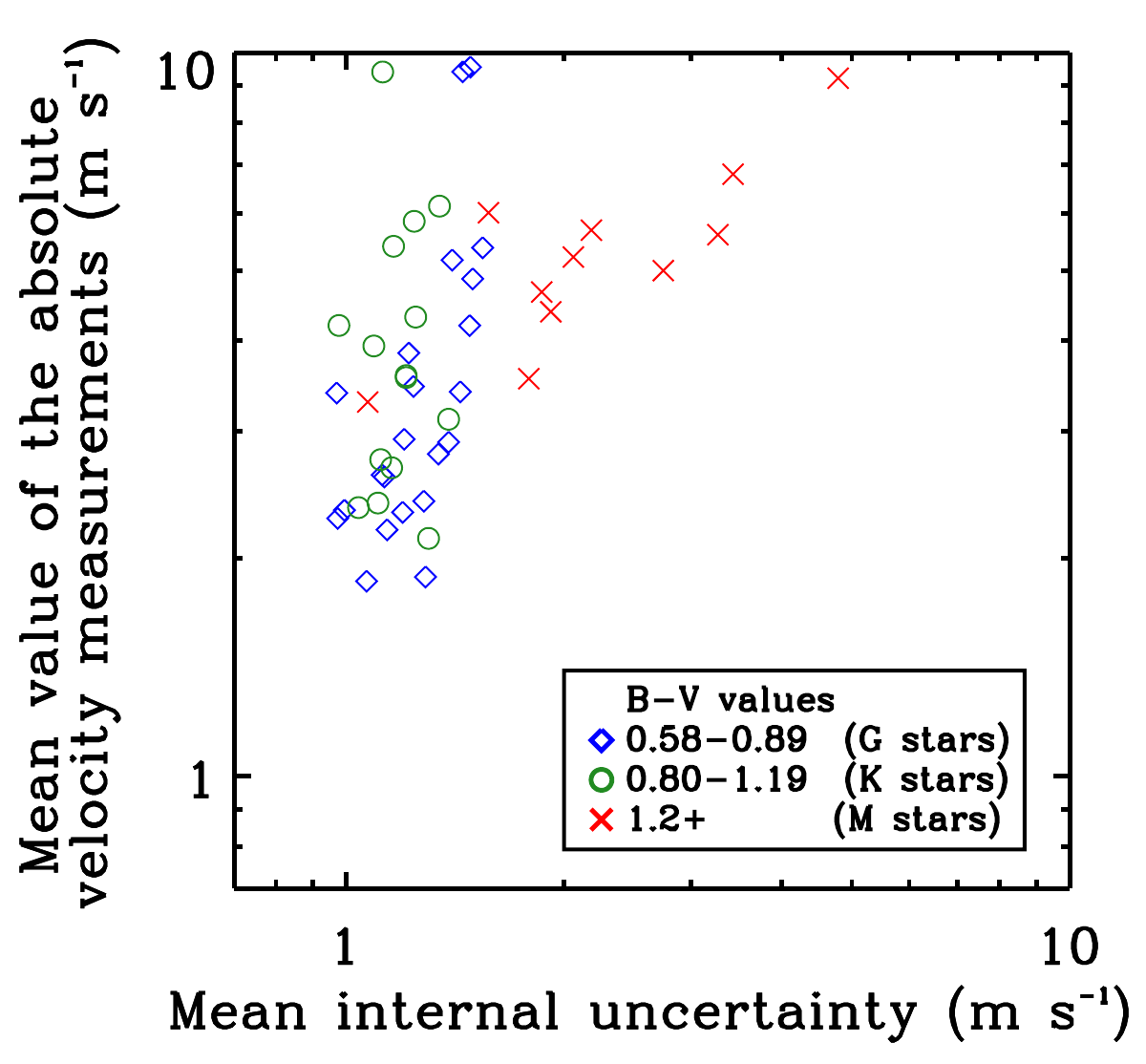}  
   \end{center}
   \caption 
   { \label{fig:muabs_meanintunc} Mean values of $\mu_{int}$ and $\mu_{abs}$  for each star observed during the same time span as the HD 185144 analysis presented in Figure \ref{fig:185144_unc}. The $\mu_{abs}$ values are always larger because they include the internal uncertainty {\it in addition} to other effects such as stellar jitter and instrument systematics. For bright, quiet stars it is possible to reach $\mu_{abs}$ values of 1.8 \ms.} 
   \end{figure}

To verify that the size of the offset between $\mu_{int}$ and $\mu_{abs}$ is not specific to the HD 185144 data set, we compare the mean values of these parameters for every star observed by the APF during the time span of Figure \ref{fig:185144_unc}. That is, we apply the same procedure detailed above for HD 185144 to all stars observed during the same date range, and then compute and compare the mean values of $\mu_{int}$ and $\mu_{abs}$ for each star's set of exposures (Figure \ref{fig:muabs_meanintunc}). As expected, the values for the mean of the absolute radial velocity values are always higher than the mean internal uncertainty values because $\mu_{abs}$ includes the effects of the internal uncertainty combined with additional sources of error such as the stellar jitter and instrument systematics. Additionally some of these stars are planet hosts, and thus display even higher $\mu_{abs}$ values because of the planet's influence. However the quietest, non-planet hosting stars are able to reach $\mu_{abs}$ values of slightly less than 2 \ms, suggesting that the additional error offset value determined using HD 185144 ($\sigma_{s} = 1.5$ \ms) is appropriate.

\section{Observing inputs}
\label{sect:Inputs}

To create an efficient and scientifically informative exoplanet survey we must balance scientific interest in a range of target stars with the observing limitations presented by the weather and the telescope's physical constraints.

\subsection{Parameters of scientific interest}
\label{sect:SciParameters}

There are a number of criteria of astronomical importance. For each star, the following parameters are utilized by the scheduling software to determine whether the star is given a high rating for observation:

\begin{itemize}
\item Observing priority - a numerical rating that reflects the observers' assessment of potential Keplerian signals in the RV data. 
\item Desired cadence - dictates the desired wait time between observations of a specific target. 
\item Desired precision - the average allowable internal velocity uncertainty for measurements of a specific target.
\end{itemize}

Both observing priority and cadence are determined via the observers' examination of the star's existing data set. As mentioned in Section \ref{sect:data}, we use the publicly available Systemic Console to analyze our RV data sets. The console produces a Lomb-Scargle periodogram from the selected RV data set, which displays peaks corresponding to periodic signals in the data. When a peak surpasses the analytic {\it false alarm probability} (FAP) threshold defined by this method, the observers note the period and the half-amplitude of the signal and use those values to decide upon an observing cadence and desired level of precision that will fill out the signal's phase curve quickly and with data points of the appropriate SNR. The desired precision is further refined if the observers have some knowledge of the star's stellar activity, as this sets a lower limit to the attainable precision. Details on the fitting procedures and the statistical capabilities of the Systemic Console are described in detail in Meschiari et. al, 2009. 

For each star, this information (along with other characteristics such as right ascension, declination, the V magnitude and the B-V color) is stored in an on-line Google spreadsheet accessible to team members and the telescope software. This database of target stars drives the survey design and target selection while also being easily accessed, understood and updated by observers. 

\subsection{Relating iodine region photons to the internal uncertainty of RV values}
\label{sect:relating_i2}

In perfect conditions (no clouds, no  loss of light due to seeing) all stars that are physically available and deemed in need of a new observation (based on their cadence) are simply ranked by observing priority and position in the night sky and then observed one after another, until dawn. Cloud cover and atmospheric turbulence, however, make such conditions rare. Furthermore, stars with low declinations spend only short periods at low air mass. Consequently there are a substantial number of constraints that affect the quality of an exposure. In order to maximize the scientific impact of each night's exposures, the conditions must be evaluated dynamically, and data taken only when the desired precision listed in the database is likely to be attainable.

To identify the targets that can be expected to attain their desired precision within the one hour observing time limit, the observing program must link the night-time conditions and the physical characteristics of each star with the resulting internal uncertainty over a given exposure time. To this end, we first relate the internal uncertainty of a given exposure to the number of photons that fall in the iodine (I$_{2}$) line-dense region of the spectrum (i.e. the $\lambda \sim 5000-6200$\AA\ bandpass) where our radial velocity analysis is performed. 

\begin{figure}[h]
   \begin{center}
   \includegraphics[height=10cm]{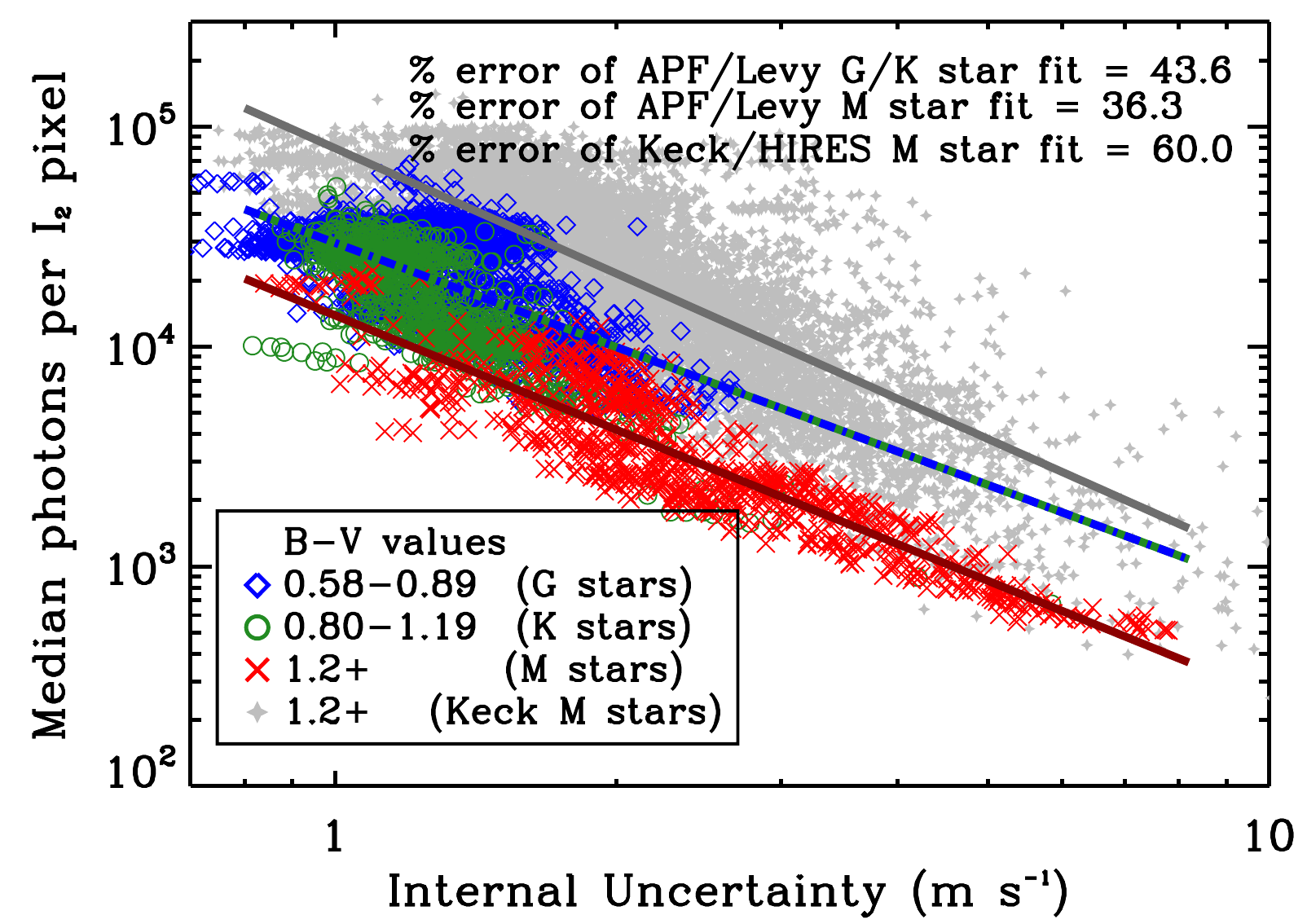}  
   \end{center}
   \caption 
   { \label{fig:I2_Uncert} Observations of G (blue), K (green) and M (red) type stars during the 1.5 years of APF/Levy observations. We plot the individual 2,790 G star exposures, 957 K star exposures and 837 M star exposures. The gray diamonds represent M star data obtained using Keck/HIRES as part of the Lick-Carnegie planet survey where the Keck/HIRES pixels have been scaled to match the same \AA\ /pixel scale as the APF/Levy ($\rm 0.0183\AA\ $ /pixel). We find that the G and K stars have the same zero point and the same slope, so we combine these two data sets for this analysis. The green and blue dashed line represents the best-fit to the APF/Levy's combined G and K star dataset, while the red line shows the best-fit to the APF/Levy M star data set which is fit separately due to the increase of spectral lines in later spectral types. As expected, the APF/Levy M stars show higher data precision for the same number of photons in the $\rm I_{2}$ region of the spectrum. The percent errors quoted on the figure are calculated using the scatter in the difference between the observed $\rm I_{2}$ photons and the $\rm I_{2}$ photons from the best-fit lines, so they represent the scatter of the sample and not the error on the mean. The dark gray line is the best-fit to the Keck/HIRES M star data set. Comparing this to the red line reveals that the APF requires 5.75x fewer photons in the $\rm I_{2}$ region to achieve velocity precision comparable to Keck/HIRES on M stars down to at least M$\rm_{v}$=10.} 
   \end{figure}

We fit the relation between internal uncertainty and photons in the iodine region separately for the G and K star data set (comprised of 2790 G star exposures and 957 K star exposures) and the M star data set (comprised of 837 M star exposures). Because information on a star's RV value comes from the location of its spectral lines, we expect the M stars - which contain many more spectral lines - to achieve better precision for the same number of photons. This expectation is validated by Figure \ref{fig:I2_Uncert} where a vertical offset between the G/K star best-fit line and the M star best-fit line is evident. Comparing the zero points for the best-fit lines (Table 1), we find a factor of 2 difference between the stellar groups. Thus the M stars reach the same level of internal uncertainty with half the number of photons required by the G and K star population.  Additionally, the slopes of both best-fit lines are close to the {\it m}=\ -2 expected for shot-noise limited observations with {\it m}=\ -1.58 and -1.73 for the G/K and M stars, respectively. The fact that the slopes are shallower than {\it m}=\ -2, indicating performance better than shot-noise limited, is reasonable as the internal uncertainty only accounts for the errors resulting from the extraction of the spectrum from the original FITS files. We emphasize that this is just a piece of the error budget, and does not include other random or systematic errors.

The functional form of these best-fit lines is given by Equation \ref{eq:I2_Unc}. Note that the numeric values for each variable (in the case of Eq \ref{eq:I2_Unc}, A and B) are listed in Table 1. This format will be used for all relations presented in \S\,\ref{sect:Inputs}.

Functional form of the fits applied in Figure \ref{fig:I2_Uncert} :

	\begin{equation}
	\label{eq:I2_Unc}
\log_{10}(N_{med}) = A + B \cdot \log_{10}(\sigma_{int}) ,
	\end{equation}

The gray points in Figure \ref{fig:I2_Uncert} correspond to spectra of M stars obtained with Keck/HIRES since November 2002 as part of the Lick-Carnegie planet search. There are 168 stars represented, all with B-V $>$ 1.2, resulting in 8872 individual exposures. In order to compare these individual velocities to those obtained on the APF in a meaningful way, we rescale the Keck/HIRES pixels so that they represent the same range of \AA\ per pixel as those on the the APF/Levy.  This involves two different scaling factors as the HIRES instrument underwent a detector upgrade in 2004 that changed its pixel size from 24$\rm \mu m$ to 15$\rm \mu m$, resulting in different sampling values. Applying these factors means that all of the data shown in Figure \ref{fig:I2_Uncert} represents the median number of $\rm e^{-}$ per pixel, where each pixel covers $\rm 0.0183\AA\ $ - the native value for the APF/Levy in the $\rm I_{2}$ region.

The figure shows that for M stars, the APF requires $\sim$5.75x fewer photons in the $I_{2}$ region to achieve velocity precision comparable to Keck\footnote{Based on the work of Bouchy et al.\ (2001)\cite{Bouchy2001}, we expect that for K dwarfs the relative speed should scale as the "information content" $Q$\cite{Connes1985}, which is proportional to the ratio of the resolutions, $Q \propto (R_{APF}/R_{HIRES})$. For HIRES, the "throughput" (the resolving power times the angular size of the slit) is 39,000" \cite{Vogt1994} and for the APF it is 114,000"\cite{Vogt2014}. Normally HIRES was used with the 0.861" slit giving a filled aperture resolution of 45k while for the Levy a 1" slit is used, so the ratio of the resolutions is 2.5. The Levy demonstrates a larger than expected improvement over HIRES which could be explained by the increased number of lines in the iodine region for M dwarfs. Further investigation is beyond the scope of the current paper but we plan to include such analysis in a future publication. We note that the excellent seeing at MK means that  some data were observed with a much higher effective R, up to 90k, which may explain the large scatter we see in Figure 5.} Speed estimates for the APF/Levy, carried out last year\cite{Vogt2014}, show that the telescope and instrument together are approximately 6x slower than Keck/HIRES. Combining these two effects indicates that the APF has essentially the same speed-on-sky as Keck/HIRES for precision RVs of M stars. This is not altogether unexpected, as HIRES was never specifically optimized for precision RV work. The APF's Levy spectrograph was purpose-built for high-precision, RV science and therefore features much higher spectral resolution and finer wavelength sampling than HIRES. Both of these factors, as well as the significantly higher system efficiency of the APF/Levy optical train over that of Keck/HIRES\cite{Vogt2014}, combine to make APF as fast as Keck/HIRES for precision RV work on M dwarfs, at least down to M$\rm_{v}$=10.

Where N$_{med}$ is the median number of photons per pixel in the I$_{2}$ region for a given exposure, and $\sigma_{int}$ is the estimated internal uncertainty for the resulting radial velocity value.

To assess whether the exposures of these stars represent a Gaussian distribution, we compare the standard RMS and the scatter calculated using a Tukey's biweight method around each fit. The Tukey's biweight scatter provides a more robust statistic for data drawn from a non-Gaussian distribution as it less heavily weights the outliers, which are assumed not to be part of a normal distribution \cite{Beers1990}. For the G and K star fit, the standard RMS is 0.182 while the biweight scatter is 0.186. Similarly, for the M stars, the standard RMS is 0.151 while the biweight scatter comes out to 0.155. 

In the limit of a true Gaussian distribution, these two metrics would produce the same result. Employing a bootstrap analysis of each method, we find the standard deviation of both the RMS and the biweight scatter of the G and K star sample to be 0.002. Similarly, the standard deviation of both the RMS and the biweight scatter of the M star sample is 0.003. Noting the similarity of these standard deviations with the actual offsets found between the RMS and biweight scatter, we determine that the observations for both sets of stars are drawn from a mostly Gaussian distribution. 

\subsection{Real-time effects}
\label{sec:realtime}

\subsubsection{Data selection}
\label{sec:dataselection}

Knowing the median number of photons per pixel in the iodine region required to achieve a given level of RV precision enables us to determine the expected exposure time for a star {\it if} we know how quickly those photons accumulate. To determine this rate and the relation between the final exposure meter value and the number of photons in the  $\rm I_{2}$ region (used to set upper bounds on observing time) we study a subset of the year of APF data described in Section \ref{sect:data}. 

Cuts are applied to the main data set to select only those observations taken on clear nights and in photometric conditions, as non-photometric data will induce skew in the results. First, we select only exposures with seeing $\rm \overline{FWHM} < 2"$, which results in 935 individual observations. We then perform separate, multi-variate linear regressions on the photon accumulation rate in the $\rm I_{2}$ region of the spectrum and the photon accumulation rate for the exposure meter (Eq \ref{eq:LinRegress_I2}, Eq \ref{eq:LinRegress_Expmeter}) on all remaining data points. In each case, we calculate the variance of deviations from the best-fitting relation for all of the data. We also calculate the variance for all of the points on a given night. We then reject nights using the F-ratio test. Namely, if the standard deviation of a given night is more than twice the standard deviation of the population as measured by the Tukey's biweight then all observations from that night are rejected. This results in data sets of 865 exposures used in the $\rm I_{2}$ photon accumulation rate regression and 816 exposures in the exposure meter photon accumulation rate regression. Points that fall significantly below the regression line are most likely due to clouds while those falling above the line are likely due to erroneous readings from the exposure meter.

Once the nights with large variance have been removed, we repeat the regressions on the remaining points to determine the actual fits described in Section \ref{sec:regressions}. In Figures \ref{fig:LinRegress_I2} and \ref{fig:LinRegress_Expmeter} the points in color are those used to perform the linear regressions, while the points in gray are those we rejected after they were deemed non-photometric. Figure \ref{fig:I2_Expmeter} uses the same set of data as Figure \ref{fig:LinRegress_Expmeter} in order to keep only the normally distributed exposure meter readings.

\subsubsection{Linear regressions}
\label{sec:regressions}

\begin{figure}[h]
   \begin{center}
   \includegraphics[height=8cm]{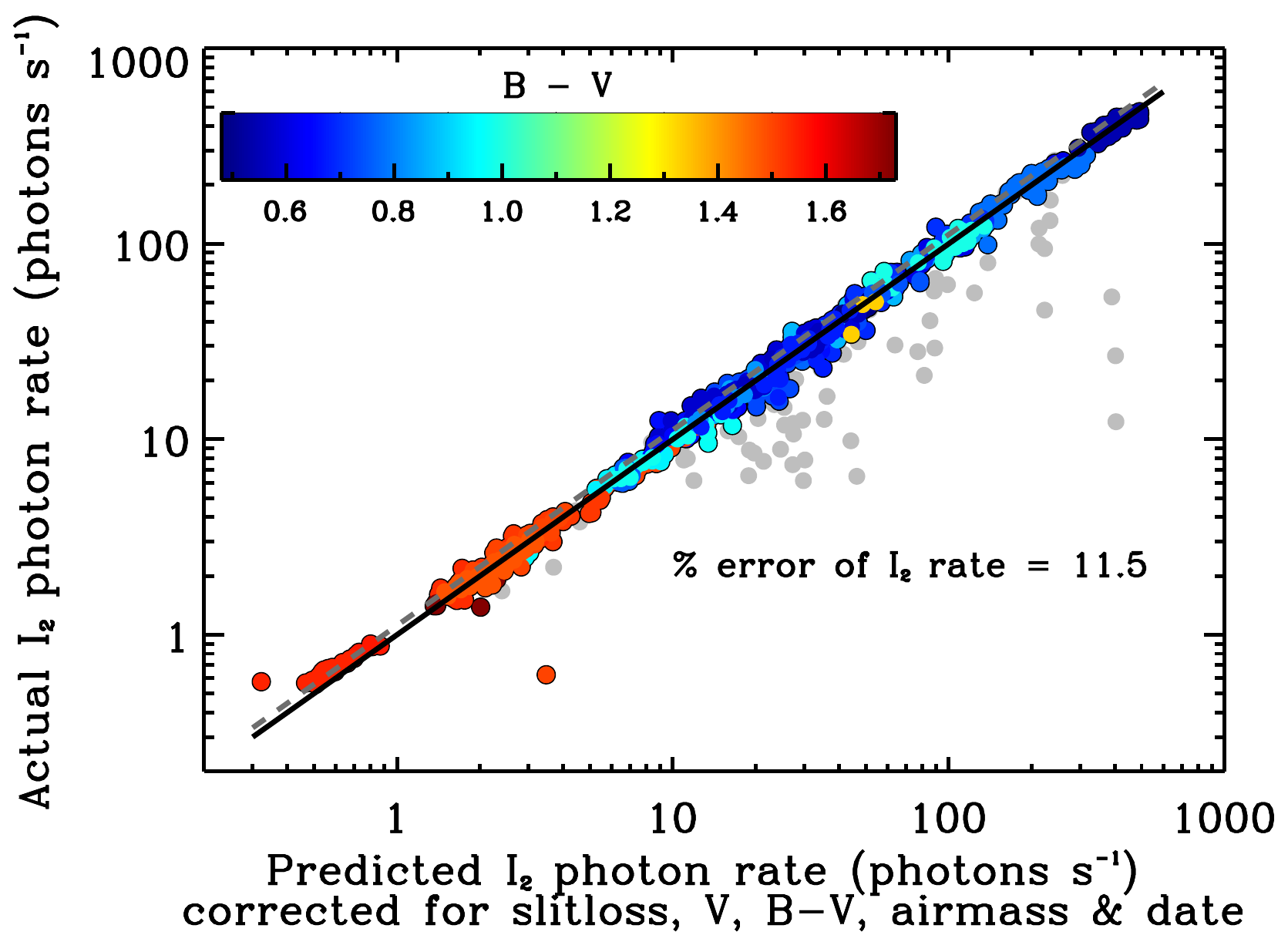}  
   \end{center}
   \caption 
   { \label{fig:LinRegress_I2} Multi-variate linear regression of the iodine pixel photon accumulation rate which incorporates stellar color, stellar magnitude, atmospheric seeing and airmass. Colored points are used in calculating the regression, while gray points have been rejected as non-photometric data as described in Section \ref{sec:dataselection}. The black line is a 1:1 relationship, and the grey dashed line shows the relation offset by one standard deviation, which is the limit we use operationally. The strong correlation between the data and the regression line enables prediction of the rate of photon accumulation in the spectrum's iodine region (a value not calculated until the data reduction process) using the stellar properties and ambient conditions. We can thus estimate the observation time required to meet a specific median I2 photon value, and, in conjunction with Figure \ref{fig:I2_Uncert}, a radial velocity precision.} 
   \end{figure}

To determine the predicted observation time of a star, we perform a multi-variate linear regression using the trimmed dataset resulting from the procedure described in the previous section. The regression estimates the rate at which photons accumulate in the pixels of the iodine region of the spectrum, and accounts for [1] the star's V magnitude, [2] its B-V color, [3] slit loss due to the current seeing conditions, [4] the airmass based on the star's location and [5] the modified date of the observation (Figure \ref{fig:LinRegress_I2}). The modified date is calculated by subtracting the maximum date from each observation following the selection process described above. This makes the zero point of the relationship the value at the time of the last photometric observation. We use the modified date parameter to address the degradation of the telescope's mirror coatings over time. When the mirrors are recoated it will introduce a discontinuity in this parameter, and we will then adjust the zero point of all regression fits based on the new throughput estimates and watch for any changes that develop in the slope of the regressions. 
 
\begin{figure}[h]
   \begin{center}
   \includegraphics[height=8cm]{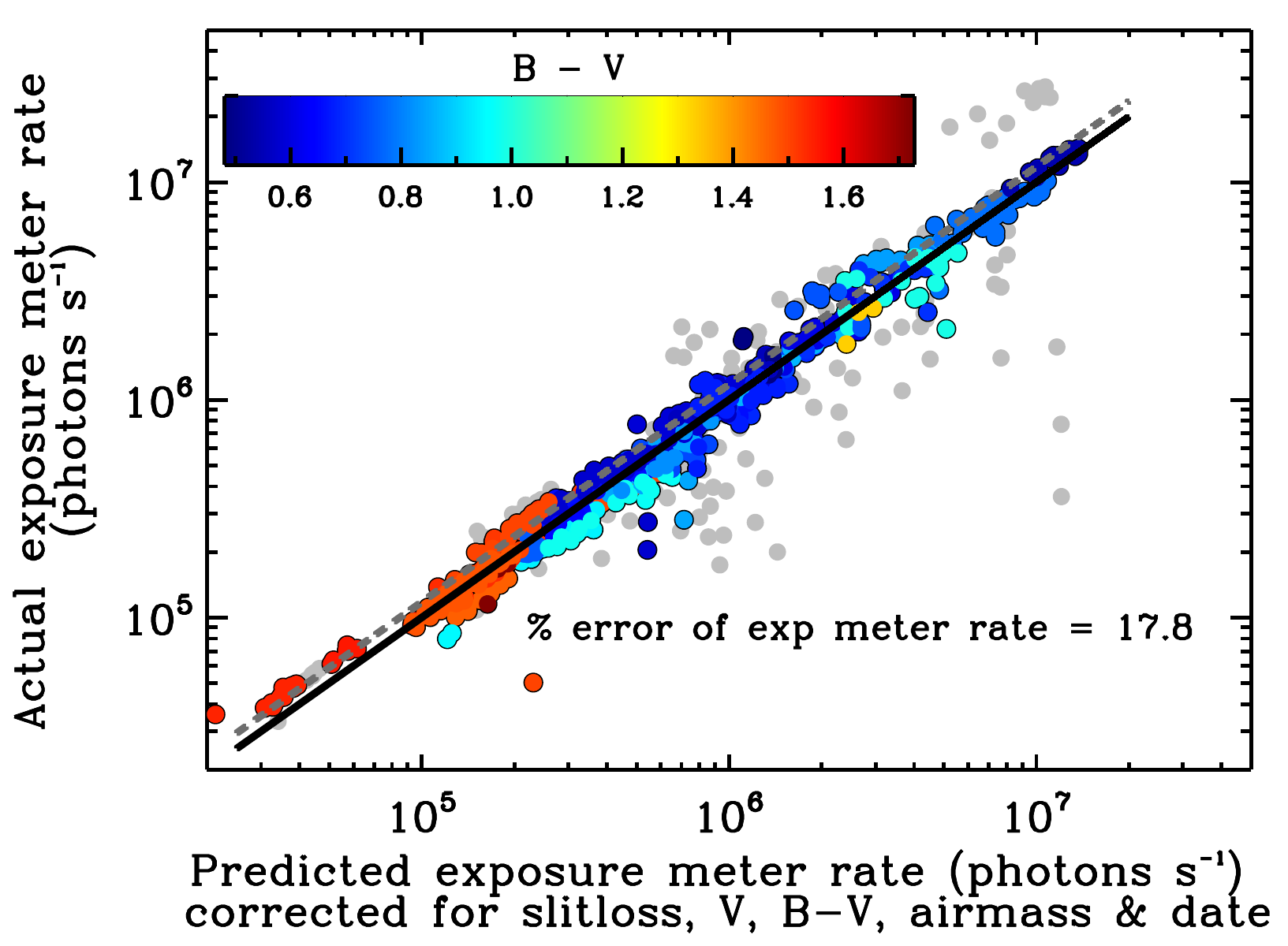}  
   \end{center}
   \caption 
   { \label{fig:LinRegress_Expmeter} Multi-variate linear regression to the exposure meter photon accumulation rate, as measured on the APF guider, which incorporates stellar color, stellar magnitude, atmospheric seeing and airmass. Colored points are used in calculating the regression, while gray points have been rejected as described in Section \ref{sec:dataselection}. The black line is a 1:1 relation, and the grey dashed line shows the relation offset by one standard deviation, which is the limit we use operationally. The strong correlation permits prediction of the expected exposure photon accumulation rate for a given star in photometric conditions, and thus provides a measure of the transparency. Any decrease in exposure photon accumulation rate from what is predicted is presumed to arise from decreases in atmospheric transparency brought about by cloud cover.} 
   \end{figure}

 The multi-parameter fit over these five variables results in a best-fit plane, of which we present a projection in Figure \ref{fig:LinRegress_I2}.  To help visualize the goodness of fit, we plot the data on one axis, the linear regression combination on the other and place a 1:1 line on top. This approach is also used when plotting the linear regression in Figure \ref{fig:LinRegress_Expmeter}. 

The regression gives :

	\begin{equation}
	\label{eq:LinRegress_I2}
\log_{10}(r_{I_{2}}) = -\frac{1}{2.5}\bigg(V_{c}+\alpha_{I}(B-V)+\beta_{I}(\sec(z))+\gamma_{I}(\mathrm{MJD})+C_{I}\bigg) ,
	\end{equation}

with

	\begin{equation}
	\label{eq:V_corr}
V_{c}=V_{m} - 2.5\cdot \log_{10}(f_{t}) ,
	\end{equation}

where r$_{I_{2}}$ is the photon accumulation rate in the iodine region, {\it z} is the angle of the star relative to zenith, MJD is a modified Julian date and $f_{t}$ is the fraction of the starlight that traverses the spectrograph slit.

By dividing the number of $I_{2}$ region photons necessary to meet our desired RV precision (derived from Eqn \ref{eq:I2_Unc}) with the photon accumulation rate in the $I_{2}$ region calculated using Eqn \ref{eq:LinRegress_I2}, we can determine the predicted observation time for a star for a specified internal uncertainty in a given set of conditions. These predicted observation times account for atmospheric conditions such as seeing and airmass, but do not address the issue of atmospheric transparency. The APF lacks an all-sky camera with sufficient sensitivity to assess the brightness of individual stars, meaning that we cannot evaluate the relative instantaneous transparency of different regions of the sky. Instead we must determine the transparency during each individual observation by comparing the rate of photon accumulation we observe with what is expected for ideal transparency.  Although the $\rm I_{2}$ region photons provide a straightforward way to determine the predicted exposure times, the number of iodine region photons is available only after the final FITS file for an observation has been reduced to yield a radial velocity measurement. Thus we cannot monitor in real time the rate at which they are registered by the detector to assess the cloud cover.

Instead, we compute a transparency estimate during each exposure using the telescope's exposure meter. The exposure meter is created by using series of 2D images from the guider camera that are updated every 1-30 seconds depending on the brightness of the target. Rather than guiding on light reflected off a mirrored slit aperture, as is traditionally done, the APF uses a beamsplitter to provide 4\% of incoming light to the guider camera as a fully symmetric, unvignetted seeing disk. This allows a straightforward way to monitor how well the telescope is tracking its target and provide realtime corrections to both under and over guiding - both of which smear out the telescope's point spread function on the CCD and result in broader full width at half maximum values for spectral lines. Guide cameras that utilize the reflected light off of a mirrored slit aperture are significantly more sensitive to these problems in good seeing, as the majority of the light falls through the spectrograph slit. In our experience, the loss of 4\% of the star's light is acceptable if it ensures that the telescope's guiding is steady throughout the night and across the different regions of the sky.

After each guiding exposure is completed, the guide camera then passes the 2D FITS images it creates to the SourceExtractor software \cite{Bertin1996} which analyzes each image and provides statistics on parameters such as the flux and full-width at half-maximum (FWHM), which are in turn used to evaluate the current atmospheric seeing. These guide camera images are also used to meter the exposures. Each image is integrated over the rectangular aperture corresponding to the utilized spectrograph slit, with background photons (determined using adjacent, background-estimating rectangles) subtracted off to determine the number of star-generated photons accumulated by the guide camera.

Analysis of the existing APF data suggests that the exposure meter rate (much like the iodine photon accumulation rate) depends on the star's color, its V magnitude, the atmospheric seeing (in the form of slit losses), the airmass and the date of observation. A multi-variate linear regression to the exposure meter rate over these five terms results in the correlation displayed in Figure \ref{fig:LinRegress_Expmeter}. 

Because the exposure meter is rapidly updated, we can monitor photon accumulation in real-time during an observation. Atmospheric seeing is already incorporated into the $V_{c}$ term in the linear regression (Eqns \ref{eq:V_corr}, \ref{eq:LinRegress_Expmeter}), so any decrease from the expected exposure meter rate likely stems from an increase in clouds and corresponding decrease in atmospheric transparency. The ratio of expected exposure meter rate to observed exposure meter rate provides a ``slowdown" factor that the scheduler tracks throughout the night and multiplies by the predicted `clear night' observation times calculated using Eqn \ref{eq:LinRegress_I2} to determine a best guess exposure duration.

The regression in Figure \ref{fig:LinRegress_Expmeter} gives :

	\begin{equation}
	\label{eq:LinRegress_Expmeter}
\log_{10}(r_{E}) = -\frac{1}{2.5}\bigg(V_{c}+\alpha_{E}(B-V)+\beta_{E}(\sec(z))+\gamma_{E}(\mathrm{MJD})+C_{E}\bigg) ,
	\end{equation}

where r$_{E}$ is the photon accumulation rate on the exposure meter, $\alpha_{E}$, $\beta_{E}$, $\gamma_{E}$ and C$_{E}$ have the same meanings as in Eqn \ref{eq:LinRegress_I2} but for the exposure meter photon accumulation rate instead of the Iodine photon accumulation rate, {\it z} is the star's zenith angle, and V$_{c}$ is defined in Eqn  \ref{eq:V_corr}.

\subsection{Setting upper bounds on exposure time}
\label{sect:upperbounds}

Finally, because of the scatter seen in Figures \ref{fig:LinRegress_I2} and \ref{fig:LinRegress_Expmeter}, we must ensure that exposures end when photons sufficient to achieve the desired RV precision have accumulated instead of continuing on for extra seconds or minutes. Photons in the $\rm I_{2}$ region can't be monitored during the observation. However the APF data set shows a strong relationship between the number of photons obtained in the $\rm I_{2}$ region of the spectrum and the photons registered by the exposure meter, which is monitored in realtime. 

\begin{figure}[h]
   \begin{center}
   \includegraphics[height=8cm]{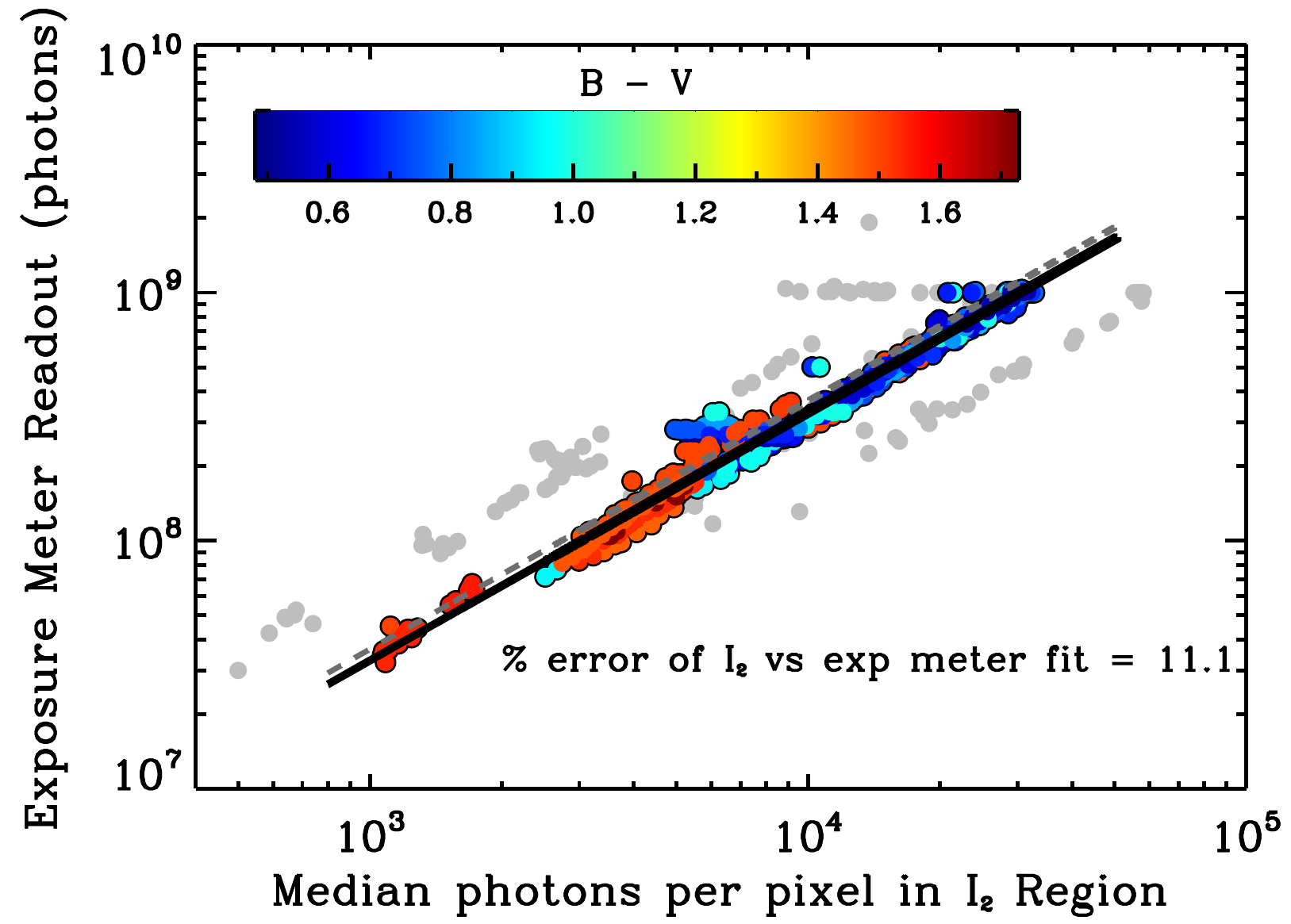}  
   \end{center}
   \caption 
   { \label{fig:I2_Expmeter} Color-corrected relationship between the photons in the $\rm I_{2}$ region of the spectrum and the photons registered by the exposure meter. The black line shows the best-fit, and the grey dashed line shows the relation offset by one standard deviation, which is the limit we use operationally. To increase telescope efficiency, we require a way to ensure observations don't continue when the number of iodine region photons necessary to achieve the desired internal uncertainty has already been achieved. As described in Section \ref{sec:regressions} there is no way to measure the $\rm I_{2}$ region photon accumulation in real time. However, the tight correlation between $\rm I_{2}$ photons and photons on the exposure meter (which does update in real time) displayed here allows us to set a maximum exposure meter value based on our desired precision level. Thus the observation software will end the exposure when the specified exposure meter value is met, even if the open shutter time falls short of the predicted observation time. This is particularly useful for cases where the cloud cover used in calculating the predicted observation time is actually more than the cloud cover on the target, which would result in an erroneously long observation.} 
   \end{figure}

The telescope's guide camera, which is used for the exposure meter, has a broad bandpass and is unfiltered. This generates a strong color-dependent bias when comparing the guider photons to those that fall in the much narrower $\rm I_{2}$ region. We apply a quadratic B-V color correction term to produce the relation shown in Figure \ref{fig:I2_Expmeter}. Combining this with the equations identified in Figure \ref{fig:I2_Uncert}, we obtain relations that allow us to relate the desired internal precision to the corresponding number of photons in the iodine region of the spectrum, and then to the number of photons required on the exposure meter. The resulting exposure meter threshold is used to place an upper limit on the exposure. This is particularly useful on nights with patchy clouds, where the cloud cover estimate calculated during the previous observation can be significantly higher than the cloud cover in other parts of the sky - resulting in artificially high predicted observation times. In this case, the exposure meter can be used to stop an observation if the desired photon count is reached early, improving efficiency.

Fit applied in Figure \ref{fig:I2_Expmeter} :

	\begin{equation}
	\label{eq:I2_Expmeter}
\log_{10}(R) = \delta + \epsilon \rm (B-V) + \zeta \rm (B-V)^{2} ,
	\end{equation}

where R is the ratio of photons on the exposure meter to photons in the iodine region of the spectrum.

\subsection{Combining the fits}
\label{sect:combining}

The above relations are combined to enable the scheduling algorithm to select scientifically optimal targets. The following list summarizes the combination scheme.
\\

{\bf Steps to determining an object's predicted observation time and exposure meter cut-off}
\begin{enumerate}
\item Query spreadsheet for stellar attributes (V, B-V, required precision, RA, Dec, observing priority).
\item Use Equation \ref{eq:I2_Unc} to determine the desired number of photons in the $I_{2}$ region of the spectrum. 
\item Use Equation \ref{eq:LinRegress_I2} to calculate the observation time required to obtain the desired total number of $\rm I_{2}$ region photons in ideal transparency conditions.
\item Multiply the slowdown factor, calculated during the previous observation using Eqn \ref{eq:LinRegress_Expmeter}, with the ideal transparency observation time estimates to get the predicted observation time in current conditions.
\item Calculate the exposure meter threshold based on the required number of $I_{2}$ region photons using Eqn \ref{eq:I2_Expmeter}.
\end{enumerate}

We find scatters of 11.5\% and 11.1\% in Figures \ref{fig:LinRegress_I2} and \ref{fig:I2_Expmeter} respectively (Table 1). This means that, even on a photometric night, we may not accumulate the number of photons in the $\rm I_{2}$ region necessary for the desired precision as the photon arrival rate could be too low. To increase the likelihood of getting enough photons to reach the desired number of photons in the $\rm I_{2}$ region, we increase the observation time estimate and the exposure meter threshold by 11.5\% and 11.1\%, or one standard deviation. By implementing this padding factor we ensure that 84\% of the time we observe a target, we will obtain the desired number of $I_{2}$ photons. However, this does not necessarily guarantee that we will achieve the desired internal uncertainty, due to the scatter in the relation between the $\rm I_{2}$ region photons and the uncertainty estimates seen in Figure \ref{fig:I2_Uncert} and quantified in Table 1.

Using these predicted observation times, the scheduler can evaluate whether any potential target can be observed at its desired precision within the one-hour observation time limit. Combining the predicted observation times with the targets' coordinates determines whether it will remain within the allowed $20-85^{\circ}$ elevation range during the exposure. Stars that satisfy all these criteria are then ranked based on their observing priority, time past cadence requirement, and distance from the moon, with the highest scoring star being selected for observation. The scheduler then transmits the necessary information for the selected star, including its expected observation time and exposure meter threshold, to the observing software, breaking up the total observation time into a number of individual exposures if necessary. 

\section{Dismissed factors}
\label{sect:DismissedInputs}

\begin{figure}[ht]
   \begin{center}
   \includegraphics[height=8cm]{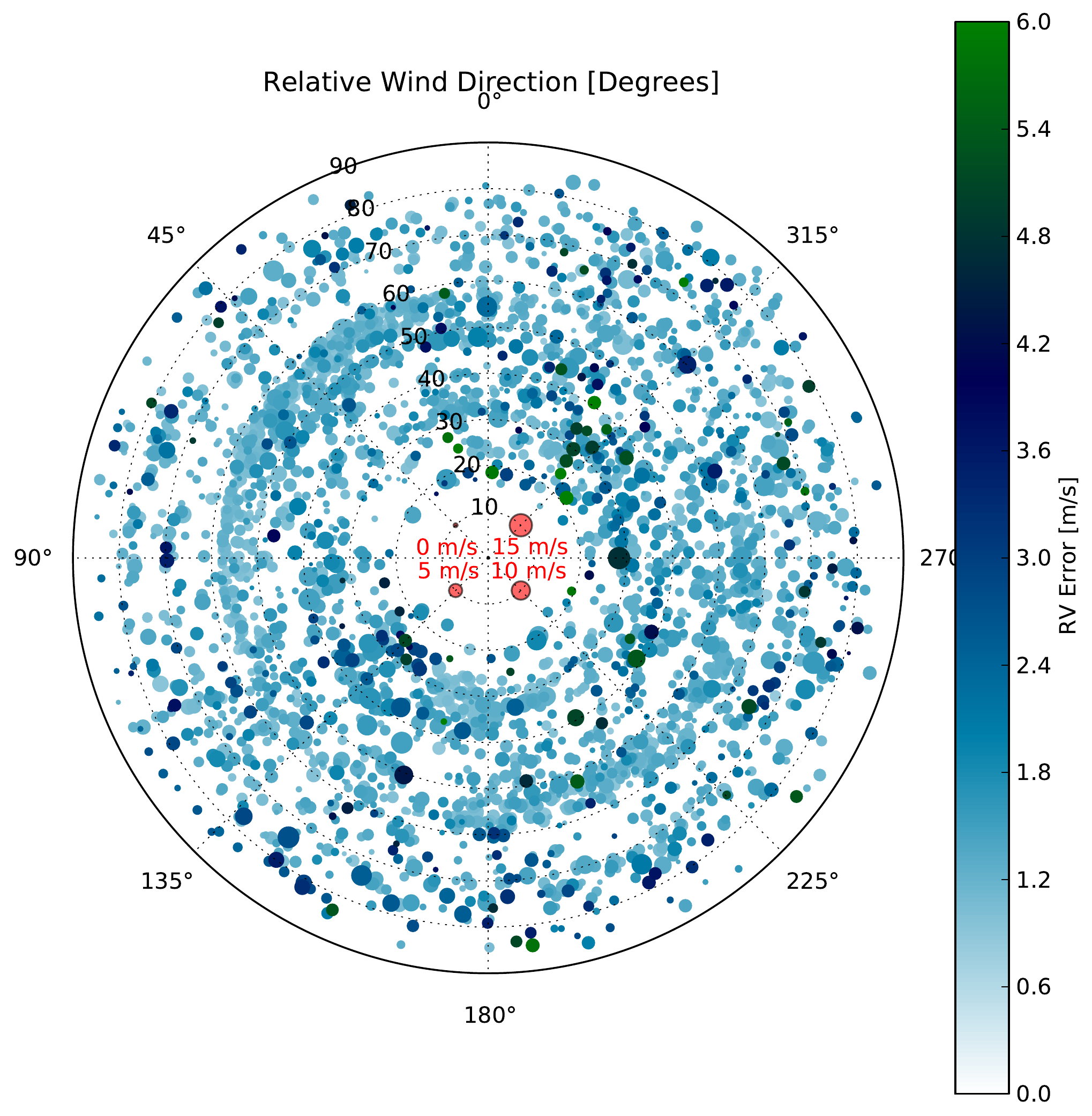}  
   \end{center}
   \caption 
   { \label{fig:Wind_plot} Windspeed (point size) and direction (azimuthal position) plotted for 3155 individual exposures reveals no strong correlation between pointing near/into the wind and the estimated internal uncertainty displayed in the color bar. The exposures represented in this figure were obtained before we had the means to determine condition-based exposure times. Thus all exposures were run until the reached their exposure meter threshold, or up to a static maximum exposure time of 900s and then terminated, regardless of the number of photons collected by the exposure meter. This means the wind based effects are not being mitigated by longer exposure times, and wind direction can thus be ignored when deciding which stars are considered viable targets for the next observation.} 
   \end{figure}

Exposures obtained during 2013 and 2014 indicate that some factors initially suspected to be important need not impact target selection considerations. For example, the original observing protocol avoided targets within 45$^{\circ}$ of the direction of any wind above 5 mph to avoid wind shake in the telescope. We find, however, no discernible increase in the internal uncertainty (indicated by the color scale in Figure \ref{fig:Wind_plot}) as a function of wind speed or direction. This resilience likely stems from an effective wind shielding mode for the dome shutters, which opens them just enough to ensure that there is no vignetting of the target star \cite{Vogt2014}. In addition, substantial effort has been put into tuning the telescope's servo motors in order to ``stiffen" the telescope and thus mitigate the effect of wind gusts that do manage to enter the dome \cite{Lanclos2014}. 

\begin{figure}[ht]
   \begin{center}
   \includegraphics[height=8cm]{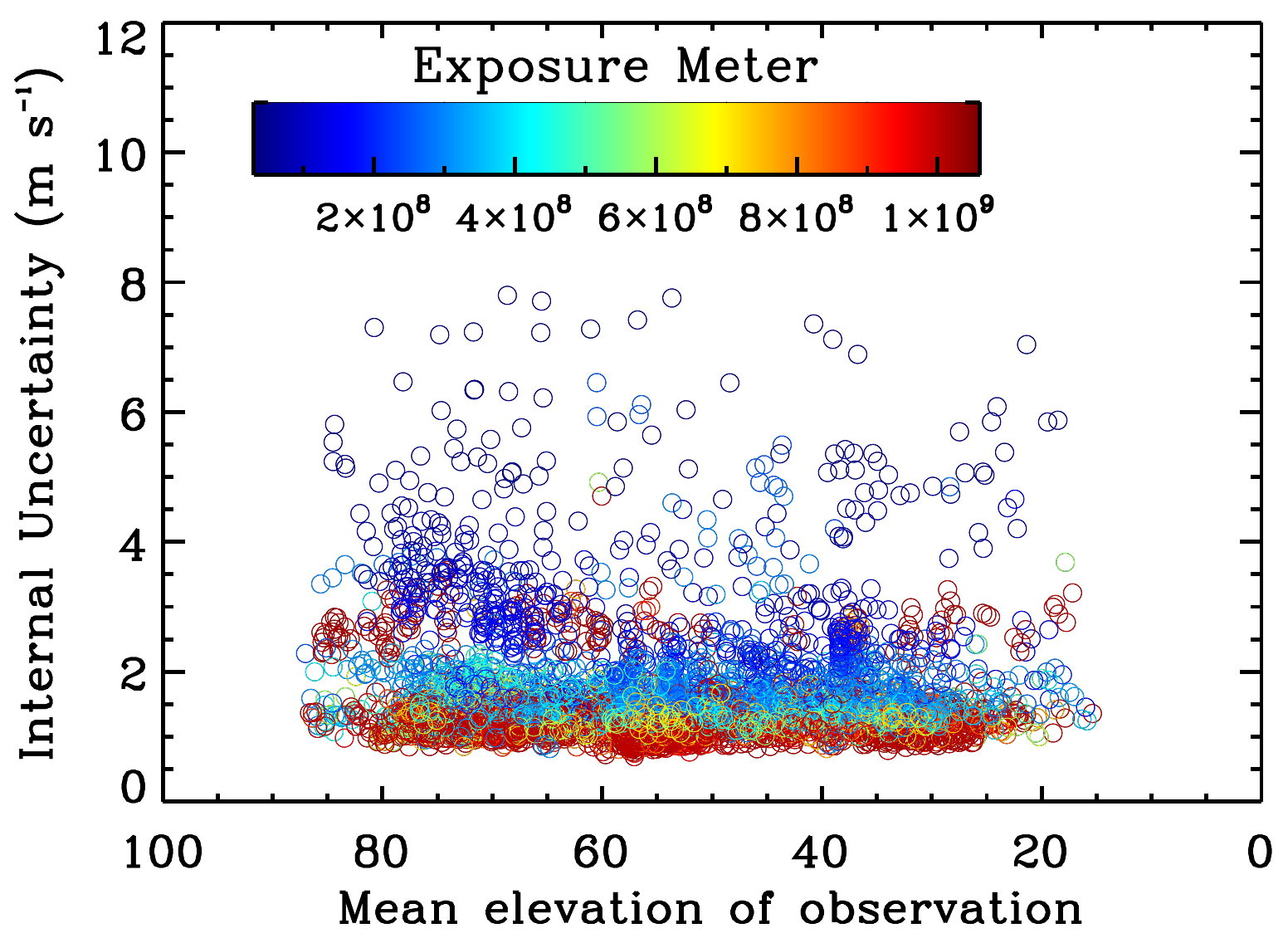}  
   \end{center}
   \caption 
   { \label{fig:El_error} The radial velocity precision as a function of the elevation shows no strong correlation, once we compare observations with a fixed exposure meter value. As our linear regressions in Figure 4 and 5 account for the decline in the photon accumulation rate with decreasing elevation, we do not need to add an additional term to account for other elevation effects such as increased seeing. Similar to the Figure \ref{fig:Wind_plot}, the observations presented here were taken before the adaptive exposure time software was in use. Thus every exposure has a static maximum exposure time of 900s and the low elevation effects are not being mitigated by allowing for longer exposures. The telescope's ADC only functions down to 15$^{\circ}$, the same elevation at which the telescope begins to vignette on the dome shutter, thus providing the lower limit for our observations.} 
   \end{figure}

We also previously assigned higher priority to targets with elevation in the 60-70$^{\circ}$ range, removing scientifically interesting targets that were closer to the horizon from consideration. As shown in Figure \ref{fig:El_error}, however, there is no significant loss in velocity precision as a function of elevation from 90$^{\circ}$ down to 20$^{\circ}$ thanks to the telescope's atmospheric dispersion corrector (ADC) which works down to 15$^{\circ}$. The telescope has a hard observing limit of 15$^{\circ}$ because of the ADC's range and because working at lower elevations leads to vignetting by the dome shutters. We still enforce an elevation range of $20-85^{\circ}$ to avoid mechanical problems in telescope tracking at the high elevations.. Working at elevations approaching our lower limit does result in longer predicted observation times (due to the airmass term) which can result in stars being skipped over in favor of other, higher elevation targets. 

Finally, we no longer assign a weighting value to the slew time necessary to move between targets. The APF is capable of moving at 3\degs\ in azimuth and 2\degs\ in elevation, which means that a direct slew to a target across the sky would take only one minute. Because the CCD takes approximately 40 seconds to read out each observation, this slew time factor is small enough to be considered unimportant. Furthermore, with the introduction of the wind shielding mode, the telescope's movement was altered so that it must first drop to a ``safe elevation" of $15^{\circ}$ before rotating to the target azimuth and then moving upwards to the appropriate elevation. This is done to protect the primary mirror from falling debris while the dome shutters are reconfigured to minimize wind-effects for the next observation  during the slew. An additional result is that all slews take approximately the same amount of time, which provides justification for discounting slew time as an input to target selection.

\section{Dynamic scheduler overview}
\label{sect:Heimdallr}

In Section \ref{sect:Inputs}, we described a method to predict observation times for targets given their precision requirements and the current atmospheric conditions. To automate the determination of these observation times and the selection of the optimal target at any time throughout the night we have implemented a dynamic scheduler (written in Python) called {\tt Heimdallr}. 

{\tt Heimdallr} runs all of the APF's target selection efforts and interfaces with pre-existing telescope control software so that, once it submits an observation request, it waits until that set of exposures is completed before reactivating. Telescope safety, system integrity, and alerts about current weather conditions are monitored by two services called {\tt apfmon} and {\tt checkapf}. Each of these software sets has the ability to override {\tt Heimdallr} if they detect conditions that pose a threat to, or represent a problem with, the telescope. This ensures that the facility's safety is always given priority. 

Additionally, while directing the night's operations, {\tt Heimdallr} uses other pre-existing utilities including {\tt openatnight}, {\tt prep-obs} and {\tt closeup} which, as their names suggest, open the facility prior to nightfall (or when night time conditions warrant), prepare the instrument and optical train for observing, and close the facility, securing the telescope when conditions warrant. The setting of the guider camera, the configuration of dome shutters and the control of telescope movement to avoid interference with the cables wrapped around the telescope base is handled by yet another utility called {\tt scriptobs}. 

\subsection{Observing description}
\label{sect:observing}

Typically, {\tt Heimdallr} initiates in the afternoon and prompts the instrument control software to focus the instrument by obtaining a dewar focus cube (series of exposures of the quartz halogen lamp taken through the Iodine cell) using the standard observation slit. Once the software determines a satisfactory value for the instrument's dewar position (via a simple linear least squares parabolic fit to the focus values), it proceeds to take all of the calibration exposures that are required by the data reduction pipeline. Upon completing these tasks, {\tt Heimdallr} then waits until dusk, at which point it consults {\tt checkapf} to ensure that there are no problematic conditions in the weather and then {\tt apfmon} to ensure facility readiness. If both systems report safe conditions, it then opens the dome, allowing the telescope to thermalize with the outside air.

{\tt Heimdallr} runs a main loop that continuously monitors a variety of keywords supplied by the telescope. At 6$^{\circ}$ twilight, the telescope is prepped for observing and begins by choosing a rapidly rotating B star from a pre-determined list. The B star has no significant spectral lines in the $\rm I_{2}$ region and serves as a focus source for the telescope's secondary mirror while also allowing the software to determine the current atmospheric conditions (as described in Section \ref{sec:regressions}).

At 9$^{\circ}$ twilight, {\tt Heimdallr} accesses the online database of potential targets and parses it to obtain all the static parameters described in Section 3.1.  It then checks the current date and time and eliminates from consideration those stars not physically available. The scheduler then employs the stars' coordinates, B-V colors and V magnitudes, combined with the seeing and atmospheric transparency determined during the previous observation, to calculate the predicted observation time for each target given their desired precision levels. Stars unable to reach the desired precision within the one hour maximum observation time are eliminated from the potential target list.

The remaining stars are ranked based on scientific priority and time past observing cadence. The star with the highest score is  passed to {\tt scriptobs} to initiate  the exposure(s).  When the exposures finish, {\tt Heimdallr} updates the star's date of last observation in the database with the photon-weighted midpoint of the last exposure and begins the selection process anew.

When the time to 9$^{\circ}$ morning twilight becomes short enough that no star will achieve its desired level of precision before the telescope must close, {\tt Heimdallr} shuts the telescope using {\tt closeup} and initiates a series of post-observing calibration exposures. Once the calibrations finish, {\tt Heimdallr} exits.

\subsection{Other operational modes}
\label{sect:modes}

In addition to making dynamic selections from the target database during the night, {\tt Heimdallr} can also be initiated in a fixed list or ranked fixed list operational mode. The fixed list mode allows observers to design a traditional starlist that {\tt Heimdallr} will move through, sending one line at a time to {\tt scriptobs}. Any observations that are not possible (due to elevation constraints) will be skipped and the scheduler will simply move to the next line.

The ranked fixed list option allows users to provide a target list that {\tt Heimdallr} parses to determine the optimal order of observations. That is, after finishing one observation from the list, the scheduler will then perform a weighting algorithm similar to what is employed by the dynamic use mode to determine which line of the target list is best observed next. {\tt Heimdallr} keeps track of all the lines it has already selected, so that they do not get initiated twice, and will re-analyze the remaining lines after each observation to select the optimal target. This option is especially useful for observing programs that have a large number of possible targets but don't place a strong emphasis on which ones are observed during a given night.

\section{Comparing with other facilities} 
\label{sect:compare}

There are a number of automated and semi-automated facilities that perform similar observations. Examples include HARPS-N, CARMENES, the Robo-AO facility at Palomar and the Las Cumbres Observatory Global Telescope (LCOGT) network. In order to place the APF's operations in context relative to these other observatories, we will briefly discuss the approaches of these other dedicated radial velocity facilities (CARMENES, HARPS-N, and the NRES addition to LCOGT) and of the more general facility with queue scheduling (Robo-AO). We will then highlight the common approaches along with discussing what is simpler for our facility as it is dedicated to a single-use. 

\subsection{RV surveys}
\label{sect:auto_rv}

Radial Velocity surveys generally require tens or hundreds of observations of the same star to detect planetary companions. Traditionally, high-precision velocities were obtained on shared-use facilities and observing time was limited. More recently, however, purpose built systems (such as HARPS) have presented the opportunity to obtain weeks or months of contiguous nights. 

The HARPS-N instrument, installed on the Telescopio Nazionale Galileo at Roque de los Muchachos Observatory in the Canary Islands, is a premier system for generating high-precision velocities. At present, it is primarily devoted to Kepler planet candidate follow up and confirmation. The system allows users to access XML standard format files that define target objects using the Short-Term-Scheduler GUI (STS) to guide the process and then assemble the objects into an observing block. When an observer initiates an observation, these blocks and their associated observing preferences are passed to the HARPS-N Sequencer software, which places all of the telescope subsystems into the appropriate states, performs the observation, and then triggers the data reduction process \cite{Cosentino2012}. Thus, while the building of target lists has been streamlined, nightly observation planning still requires attention from an astronomer or telescope technician.

The forth-coming Calar Alto CARMENES instrument is expecting first light in 2015. \\CARMENES will also be used to obtain high-precision stellar radial velocity measurements on low-mass stars. Its automated scheduling mechanism relies on a two-pronged approach: the off-line scheduler which plans observations on a weekly to nightly time scale based on target constraints that can be known in advance, and the on-line scheduler which is called during the evening if unexpected weather or mechanical situations arise and require adapting the previously calculated nightly target list \cite{Garcia-Piquer2014}. 

Finally, the LCOGT network will soon implement the Network of Robotic Echelle Spectrographs (NRES), six high resolution optical echelle spectrographs slated for operation in late 2015. Like all instruments deployed on the LCOGT network, the scheduling and observing of NRES will be autonomous. Users submit observing requests via a web interface which are then passed to an adaptive scheduler which works to balance the requests' observing windows with the hard constraints of day and night, target visibility, and any other user specified constraints (e.g. exposure time, filter, airmass). If the observation is selected by the adaptive scheduler, it creates a ``block" observation tied to a specific telescope and time. This schedule is constructed 7 days out, but rescheduling can occur during the night if one or more telescopes become unavailable due to clouds, or if new observing requests arrive or existing requests are canceled. In this event, the schedule is recalculated, and observations are reassigned among the remaining available sites \cite{Eastman2014}.

\subsection{Automated queue scheduling}
\label{sect:queue}

Robo-AO is the first fully automated laser guide star adaptive optics instrument. It employs a fully automated queue scheduling system that selects among thousands of potential targets at a time with an observation rate of $\sim20$ objects $\rm hr^{-1}$. Its queue scheduling system employs a set of XML format files which use keywords to determine the required settings and parameters for an observation. When requested, the queue system runs each of the targets through a selection process, which first eliminates those objects that cannot be observed, and then assigns a weight to the remaining targets to determine their priority in the queue at that time. The optimal target is chosen, and the scheduler passes all observation information to the robotic system and waits for a response that the observations were successfully executed. Once the response is received, the observations are marked as completed and the relevant XML files are updated \cite{Riddle2014}.

\subsection{Comparing our approach to other efforts}
\label{sect:compare_sub}

We have designed this system based on our RV observing program carried out over the past 20 years at Keck and other facilities. This experience, coupled with the preexisting software infrastructure,  has guided the development of both the dynamic scheduler software itself and our observing strategy.

Comparison to these other observing facilities and the strategies they employ emphasizes some shared design decisions. For example, the APF has a similar target selection approach to that of the Robo-AO system. Both utilize a variety of user specified criteria to eliminate those targets unable to meet the requirements and then rank the remaining targets, passing the object with the highest score to the observing software. Additionally, as is common with all of the observing efforts mentioned above, our long term strategy is driven by our science goals and is in the hands of the astronomers involved with the project.

Several differences are also notable. The first is that we lack an explicit long term scheduling component in our software. Our observing decisions are made in real time in order to address changes in the weather and observing conditions that occur on minute to hour time scales, and to maximize the science output of nights impacted by clouds or bad seeing. However, for a successful survey there must also be a longterm observing strategy for each individual target. We address this need via a desired cadence and required precision for every potential target. By incorporating the knowledge of how often each star needs to be observed and a way to assess whether the evening's conditions are amenable to achieving the desired precision level, {\tt Heimdallr} adheres to the longterm observing strategy outlined by our observing requirements and doesn't need to generate separate multi-week observing lists.

A second difference is that the final output of the scheduler is a standard star list text file, one line in length. This format has been in use in UCO-supported facilities such as Lick and Keck Observatories for more than 20 years, and thus is familiar to the user community. The file is a simple ASCII text file with key value pairs for parameters and a set of fixed fields for the object name and coordinates. This permits ready by-eye verification of the next observation if desired and allows the user to quickly construct a custom observing line that can be inserted into the night's operations if needed. Observers can furthermore easily make a separate target list in this format for observations when not using the dynamic scheduler (see Sec \ref{sect:modes})

Finally, we use a Google spreadsheet for storing target information and observing requirements, as opposed to a more machine-friendly format such as XML. Although Lick Observatory employs a firewall to sharply limit access to the APF hosts, it only operates on the incoming direction. Thus it is straightforward and non-compromising from a security standpoint for our internal computers to send a request to Google and pull the relevant data back onto the mountain machines. This approach provides team members with an accessible, easy-to-read structure that is familiar and easily exportable to a number of other formats. Google's version control allows for careful monitoring of changes made to the spreadsheet and ensures that any accidental alterations can be quickly and easily undone. The use of the Google software also allows interaction with a browser, so no custom GUI development is required. Therefore, we are taking advantage of existing software to both minimize our development effort, and make it as easy as possible to have the scientists update and maintain the core data files that control the observations.


\section{Observing campaigns on the automated planet finder} 
\label{sect:campaigns}

The APF has operated at high precision for over a year, and has demonstrated precision levels of $\rm \sigma \sim 1$ \ms on bright, quiet stars such as Sigma Draconis\cite{Vogt2014}. The telescope's slew rate permits readout-limited cycling and 80-90\% open shutter time, allowing for 50-100 Doppler measurements on clear nights.

\subsection{The Lick-Carnegie survey} 
\label{sect:LCsurvey}

{\tt Heimdallr}'s design dovetails with the need to automate the continued selective monitoring of more than 1,000 stars observed at high Doppler precision at Keck over the past 20 years \cite{Butler2015}. The APF achieves a superior level of RV precision and much-improved per-photon efficiency in comparison to Keck/HIRES for target stars with V \textless 10. As a primary user, we can employ the APF on 100+ nights per year in the service of an exoplanet detection survey.

At present, 127 stars have been prioritized for survey-mode observation with the APF. This list emphasizes stars that benefit from the telescope's more northern location, and gives preference to stars that display prior evidence of planetary signals. We adopt a default cadence of 0.5 observations per night. When a star is selected, it is observed in a set of 5-15 minute exposures with the additional constraint that the total of a night's sequential exposures (the observation time) on the star is less than an hour. Additionally, information obtained at Keck has, in some cases, permitted estimates for the stellar activity of specific target stars. In these cases, observation times can be adjusted to conform to a less stringent desired precision.

With its ability to predict observation times, {\tt Heimdallr} readily achieves efficiencies that surpass the use of fixed lists, and indeed, its performance is comparable to that of a human observer monitoring conditions throughout the night.

\subsection{TESS pre-covery survey} 
\label{sect:TESSsurvey}

{\tt Heimdallr} is readily adopted to oversee a range of observational programs, and a particularly attractive usage mode for APF arises in connection with NASA's Transiting Exoplanet Survey Satellite (TESS) mission. TESS is scheduled for launch in 2017, and is the next transit photometry planet detection mission in NASA's pipeline \cite{Ricker2014}. Transit photometry observations of potential planet signatures generally require follow-up confirmation, with RV being the most common. Currently there is a dearth of high-precision RV facilities in the northern hemisphere, and HARPS-N is heavily committed to Kepler planet candidate follow up. NASA recently announced plans to develop an instrument for the 3.5m WIYN telescope at Kitt Peak Observatory capable of extreme precision Doppler spectrography to be used for follow up of TESS planet candidates \cite{WIYN}. However such an instrument will require time for development and commissioning and thus is not a viable candidate for pre-launch observations of TESS target stars. The APF's year-round access to the bright stars near the north ecliptic pole, which will obtain the most observation time from TESS, makes it an optimal facility to conduct surveys in support of the satellite's planet detection mission.

At present, comparatively little is known about the majority of TESS's target stars. We have little advance knowledge of which stars host properly inclined, short-period transiting planets observable by the satellite. We will thus start with a default value for the observing cadence and be poised to adapt quickly should hints of planetary signatures start to emerge. Additionally, initial desired precisions (and the corresponding observation times) must rest on the suspected stellar jitter of the targets.

The availability of on-line databases to track all of the science-based criteria will be crucial for moving back and forth quickly between this project and further Lick-Carnegie follow up. Evaluations of the value of APF coverage are governed by three scientific criteria (priority, cadence and required precision), along with supporting physical characteristics (RA, Dec, V and B-V) for each target. Thus when beginning RV support for TESS, {\tt Heimdallr} can easily be instructed to reference the TESS database when determining the next stellar target (instead of the Lick-Carnegie List).

TESS observations also provide an excellent test bed for experimenting with alternate observational strategies. For example, Sinukoff et al. (2014)\cite{Sinukoff2014} stated that obtaining three 5-minute exposures of a star spaced approximately two hours apart from one another during the night results in a 10\% increase in precision over taking a singe 15 minute exposure. The TESS stars that will be monitored by the APF are all located in the north ecliptic pole region, meaning that the slew times will be almost negligible. It is thus likely that an observing mode that subdivides exposures to improve precision could be very valuable. In short, the APF is extremely well matched to the TESS Mission.

\acknowledgments 

We are pleased to acknowledge support from the NASA TESS Mission through MIT sub award \#5710003702.
We are also grateful for partial support of this work from NSF grant AST-0908870 to SSV, and NASA grant NNX13AF60G to RPB.
This material is based upon work supported by the National Aeronautics and Space Administration through the NASA Astrobiology Institute under Cooperative Agreement Notice NNH13ZDA017C issued through the Science Mission Directorate.
This research has made use of the SIMBAD database, operated at CDS, Strasbourg, France.


\bibliographystyle{spiejour}   

\end{document}